\documentstyle[epsfig,aps]{revtex}
\begin{document}
\draft
\title{The electro production of $d^*$ dibaryon}

\author{Di Qing$^{1,2}$, He-ming Sun$^1$, and Fan Wang$^1$}
\address{$^1$Department of Physics, Nanjing University, Nanjing,
210093, China}
\address{$^2$Institute of Modern Physics, Southwest Jiaotong University,
 Chengdu, 610031, China} 
\maketitle
\begin{abstract}
 $d^*$ dibaryon study is a critical test of hadron interaction models. The
electro production cross sections of $ed\rightarrow ed^*$ have been calculated
based on the meson exchange current model and the cross section around 30 degree
of 1 GeV electron in the laboratory frame is about 10 nb. The implication of
this result for the $d^*$ dibaryon search has been discussed.
\end{abstract}


\section{Introduction}
  The fundamental blocks of strong interaction have been confirmed to be
quark and gluon. Meson, baryon, nucleus, and neutron star are the observed 
samples of the strong interaction matter. Theoretically it is expected 
that there should be other samples of the strong interaction matter such 
as exotic quark-gluon systems, strangelet, strange star, and quark-gluon
plasma. Experimental search for these new strong interaction matters 
has been undertaken for two decades, even though there are some candidates
of these new strong interaction matters but nothing new has been 
established. New facilities have been under operation or under construction
for further experimental search. To help these search projects, new 
theoretical input is needed. An estimate of the $d^*$ production cross
section through the $\pi d\rightarrow\pi d^*$ aimed to use the LAMPF 
$\pi$ beam was reported in 1989\cite{gold1}. A strong production of $d^*$,
aimed to use the existed proton machine, has been reported by Wong 
\cite{wong1}. Intensive electron beam facilities in the 
GeV energy range are available. This paper reports an  electro-production 
calculation of $d^*\left(IJ^p=03^+\right)$\cite{gold1,wang1}. 

  There are experimental indications of dibaryon states. A high mass
($\sim2.7~GeV$) dibaryon was reported by the Sacley group\cite{le} and a low
mass($2.06~GeV$) one was reported by Moscow-Tuebingen group\cite{bi} in 
addition to other more tentative ones. H particle\cite{ja} has been 
hunted for more than 20 years. Why is $d^*$ interesting?

  Dibaryon states are closely related to the hadronic interaction, which 
is too complicated to be studied directly from the fundamental strong
interaction theory, the QCD at present. Many QCD inspired models have 
been developed to describe the hadronic interaction. 

The meson exchange 
model, based on meson-baryon coupling, was developed long before 
QCD\cite{yu} and it is still  best at fitting the experimental 
data quantitatively\cite{dema}. However its validity in QCD is not clear 
at the moment. Moreover there are many phenomenological parameters 
fixed in the fitting process and in turn it is hard to make definite 
predictions for new physics such as the dibaryon states\cite{kf}. 

Chiral
perturbation effective field theory\cite{wein} employs Goldstone bosons,
resulting from spontaneous chiral symmetry breaking, as the effective
degree of freedom in the low energy region, and the extension to the N-N
interaction is encouraging\cite{ka}.  Dibaryon has not been studied in
this approach yet.  

L.Ya.~Glozman, D.O.~Riska and G.E.~Brown\cite{gl} proposed that the
Goldstone boson is not only an effective degree of freedom for describing
hadronic interactions but also good for analyzing baryon internal
structure. In their model, constituent quarks and Goldstone bosons are
used as the effective degrees of freedom. Up to now the application is
mainly restricted in the baryon spectroscopy except a study on the 
origin of the N-N repulsive core. 

A.~Manohar and
H.~Georgi\cite{mg}, however, have argued that between the chiral symmetry
breaking scale($\sim1~GeV$) and the confinement scale($\sim0.2~GeV$), the
effective degrees of freedom are Goldstone bosons, constituent quarks
and gluons.  Such a hybrid quark-gluon-meson exchange model has been
developed to describe nucleon-baryon interactions and a
semi-quantitative fit has been obtained\cite{fu}. Some dibaryon states have
been studied with this model\cite{fae}. 

A constituent quark
and effective one gluon exchange model\cite{rgg} describes hadron
spectroscopy quite well\cite{is1}, but only the repulsive core of the N-N
interaction was obtained when this model was extended to study hadron
interactions\cite{wong}.  

The MIT bag model uses the current quark and
gluon to describe hadron internal structure\cite{cjj}. It  was used
extensively in the study of dibaryon states in the early 1980's 
and resulted in an explosion of dibaryon states
\cite{dsw}. It was realized latter that the unphysical boundary condition
should be modified\cite{is2}. One modified version is the R-matrix method
\cite{lo} and the other one is the compound quark model approach\cite{sim}.

The Skyrme model\cite{sw} has also been used to study hadron interactions 
\cite{wa} and dibaryons\cite{jaf}. Additional models might exist that
should be added to the list. 

It seems hard to discriminate these models just by the hadron spectroscopy 
and the existed scattering data of hadron interactions.
Theoretically it is also hard to justify which effective degrees of freedom
are the proper ones. On the other hand the well known phenomena, that the
nucleus is a collection of nucleons rather than quarks, and that the
nuclear force has similarities to the molecular force except for energy
and length scale differences, have not been explained by any of these
model approaches.

 A pure quark-gluon model description of the $N$-$N$ interaction has
been developed\cite{wang1,wang2}. It starts from a multiquark system
and demonstrates that in the $N$-$N$ channels, it is energetically
favorable for the system to cluster into two nucleons and that the
nuclear intermediate range attraction is caused by quark delocalization
similar to the electron delocalization which induces intermediate range
molecular attraction. In the 9 and 12 quark systems with quantum numbers
of $^3H$, $^3He$ and $^4He$ the nucleon clustering has been verified
as well as the two nucleon system\cite{gold2}. 
This model has been extended to $N$-$\Lambda$ and
$N$-$\Sigma$ interactions\cite{wu} and the results show that the quarks
delocalize properly in different channels to induce qualitatively
correct $N$-$N$ (JI = 10, 01, 11, 00), $N$-$\Lambda$
(JI = 1$\frac{1}{2}$, 0$\frac{1}{2}$), and $N$-$\Sigma$
(JI = 1$\frac{1}{2}$, 0$\frac{1}{2}$, 1$\frac{3}{2}$, 0$\frac{3}{2}$)
interactions. For other channels\cite{wang1}, such as $\Delta\Delta$ 
(JI = 30), $\Delta\Omega$(3$\frac{3}{2}$), $\Delta\Sigma^*$(3$\frac{1}{2}$)
 and $\Delta\Xi^*$(31), it is energetically favorable 
for the quarks to merge into quark matter instead of two baryons, and 
there are often strong effective attractions in these channels. In the
$\Delta\Delta$ (JI = 30), the $d^*$ channel, the effective attraction is so
strong that the total energy of the system($\sim 2.1~GeV$) is near the $NN\pi$
threshold; therefore the $d^*$(JI = 30) might be a narrow resonance state
\cite{wong1}. 

Different model approaches give quite different mass of $d^*$. 
The meson baryon coupling model\cite{kf} gave a $\Delta-\Delta$
binding of $11-340~MeV$ depending on the coupling constants of $\rho$ and $\omega$
and the hard core radii($0.2-0.3~fm$). If the hard core radii are larger than
$0.38-0.40 ~fm$,the binding energy will be less than $10~ MeV$. The
quark-meson-gluon hybrid model obtained a binding between $37~ MeV$ and $80~MeV$
\cite{yz}. The pure gluon exchange model got $18~ MeV$ binding even though there is a
repulsive core in many other baryon-baryon channels\cite{okay}. MIT bag
model will give a $d^*$ mass of $2340~MeV$ if the bag radius is adjusted to give
the minimum. On the other hand the R-matrix version of the modified bag
model give a mass as high as $2840~MeV$\cite{lo}. The skyrmion model obtained
a very weak binding $\sim10~MeV$\cite{wal}. Therefore $d^*$ study will provide
a critical test of hadron interaction models. (This was already pointed 
out in the US Long Range Plan for Nuclear Science in 1996\cite{nsf}.)
 
 Quark delocalization plays a vital role in lowering the $d^*$ mass. In a
variational calculation this is nothing else but just a method to enlarge
the variational Hilbert space. Therefore it might be a general property of
quantum mechanics. If it is really realized in the nucleus(nucleon swollen explaination of the EMC effect
\cite{cj} might be taken as an evidence of quark delocalization),
it has far-reaching implications however. 
The phase transition from nuclear matter to quark-gluon plasma would be at
best a second order one or just a crossover as happened in the transition from atomic gases into plasma and would make the already hard identification of
this phase transition even harder\cite{wong1}.  $d^*$ search is a critical test
of the quark delocalization mechanism.

\section{Electo-production mechanism of $d^*$}
 $d^*$ is a spin 3 state. Its dominant hadronic component is $\Delta\Delta$.
To produce a $d^*$ from a nuclear target, one has to change two nucleons into
two $\Delta$'s. A virtual photon exchange can only excite one nucleon into
a $\Delta$. Double photon exchange is a fourth order QED process. Its
contribution to the cross section will be proportional to $\alpha^4$, a too
small effect. Here $\alpha=e^2/\left(\hbar c\right)$. Therefore the electro-
production of $d^*$ is critically dependent on the gluon exchange current
in the quark-gluon description and on the meson exchange current in the
hadronic description. The gluon exchange current is unique as depicted in 
Fig.1. 

\begin{figure}
\begin{center}
\rotatebox{180}{\psfig{figure=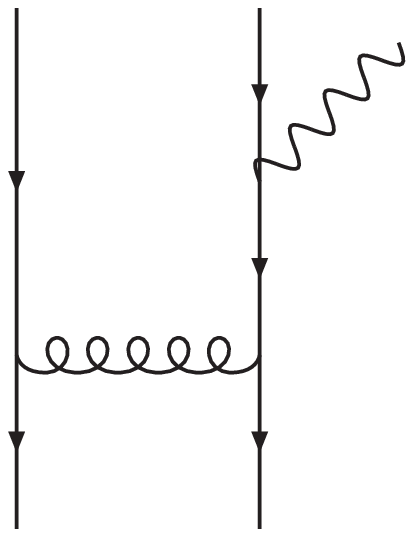,height=6cm}}
\caption{Quark gluon exchange current}

\rotatebox{180}{\psfig{figure=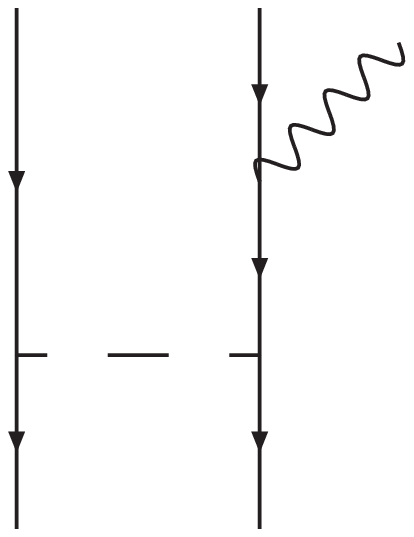,height=6cm}}
\caption{Quark meson exchange current}
\end{center}
\end{figure}

However as mentioned before that some models emphasized the Goldstone boson
quark coupling. Then there will be meson exchange current as depicted in
Fig.2 even within a baryon. This makes the exchange current in the quark-
gluon description rather model dependent. J.A.Gomez Tejedor and E. Oset(GO) 
calculated the $ed\rightarrow e^\prime p\Delta$ and $\gamma d\rightarrow$
$\Delta\Delta\rightarrow pn \pi^+\pi^-$ cross sections with the hadronic
degree of freedom\cite{go}. This approach is complicated by the
many effective meson-baryon couplings. However those coupling constants
have been fixed by the experimental data. We will take GO approach to do
the $d^*$ electro-production cross section calculation. Based on the
isospin and angular momentum conservation and taking into account the results of GO, the
following Feynman diagrams are included in our calculation,

\begin{figure}
\begin{center}
\rotatebox{180}{\psfig{figure=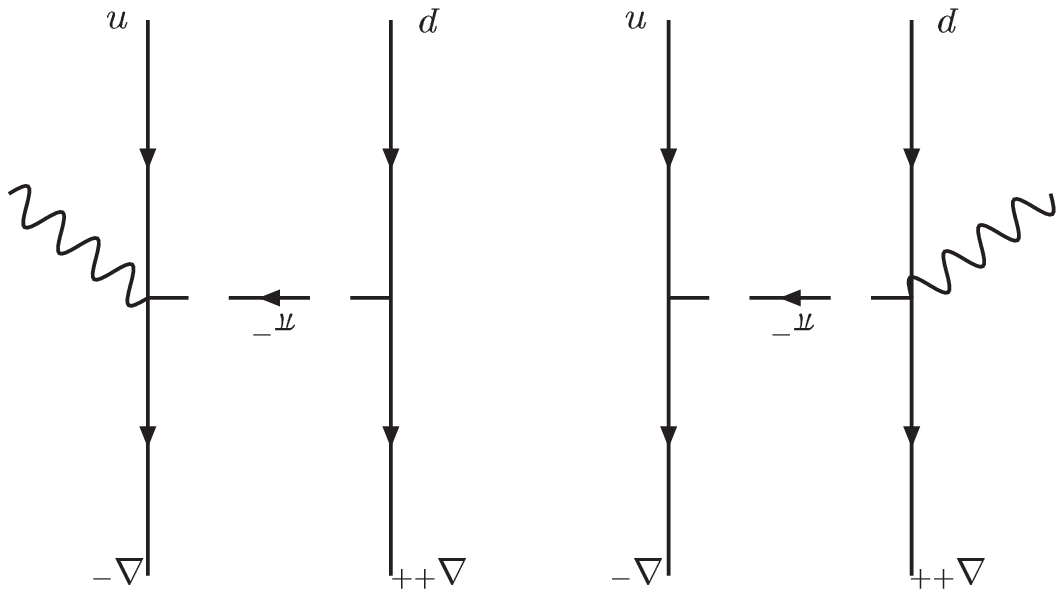,height=6cm}}
\caption{Kroll-Ruderman process}

\rotatebox{180}{\psfig{figure=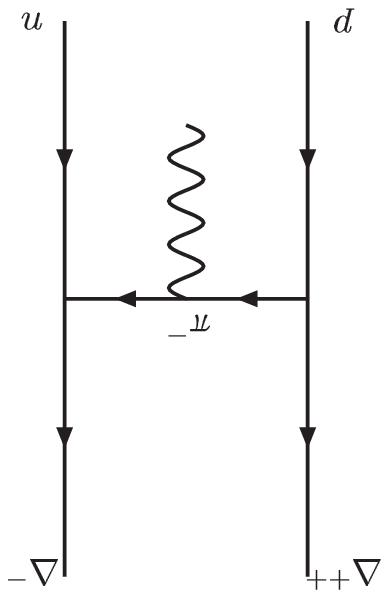,height=6cm}}
\caption{Meson exchange current}
\end{center}
\end{figure}

\begin{figure}
\begin{center}
\rotatebox{180}{\psfig{figure=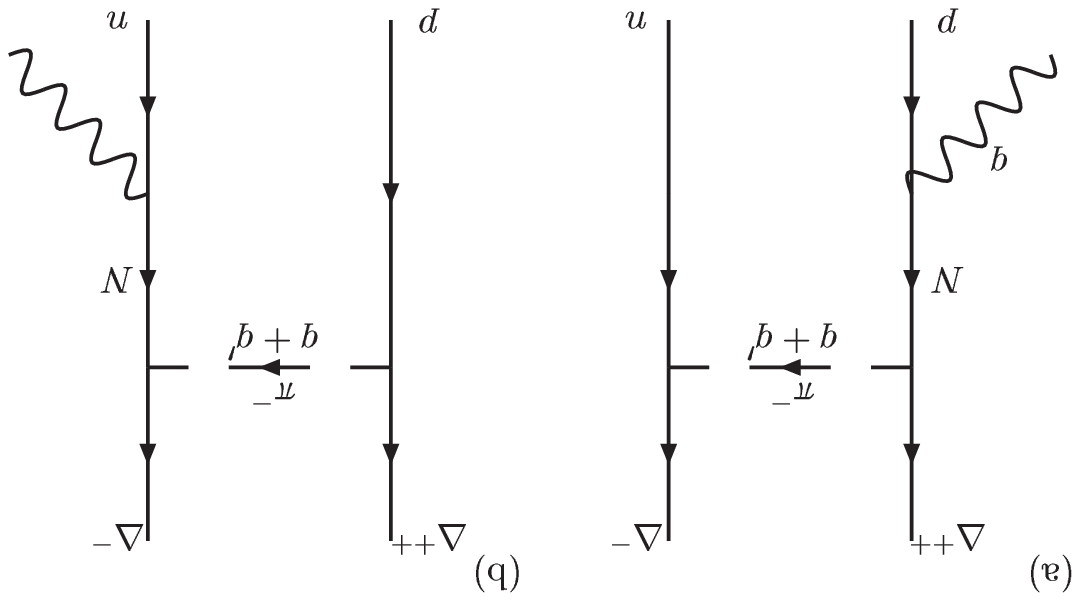,height=6cm}}
\caption{$NN$ intermediate state}
\end{center}
\end{figure}

\begin{figure}
\begin{center}
\rotatebox{180}{\psfig{figure=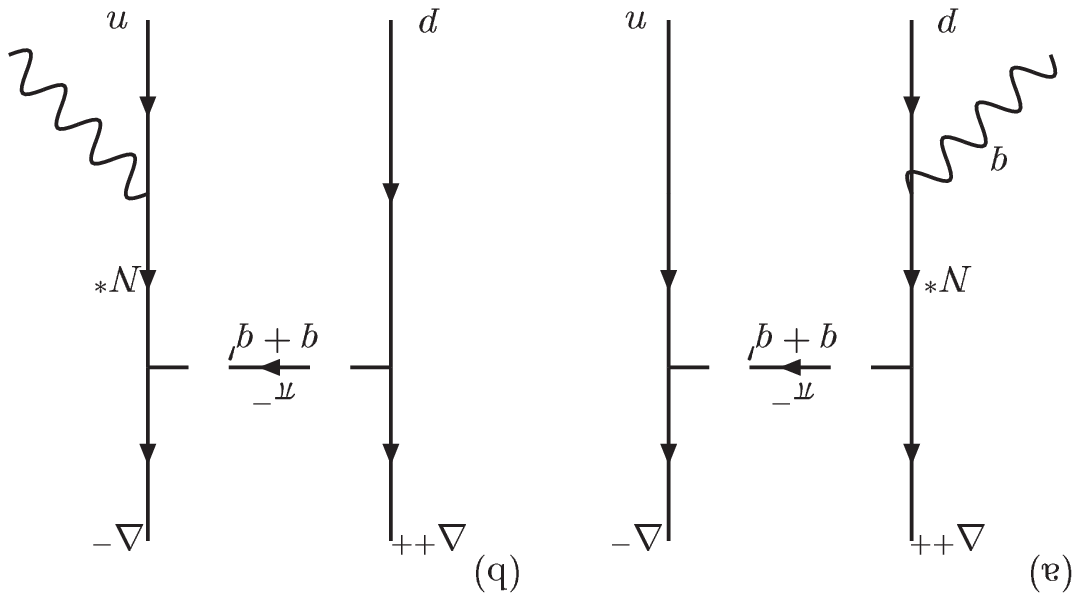,height=6cm}}
\caption{$NN^*$ intermediate state}
\end{center}
\end{figure}

The $d^*$ is an isospin I=0 state,
\begin{eqnarray} 
\mid d^*\rangle&=&\frac{1}{2}(\Delta^{++}\Delta^--\Delta^+\Delta^\circ
+\Delta^\circ\Delta^+-\Delta^-\Delta^{++}).
\end{eqnarray}
The Kroll-Ruderman term Fig.3 does not contribute due to the cancellation 
of the $\pi^+$ and $\pi^-$ exchange between different components of $d^*$
and d.
The meson exchange term Fig.4. does not contribute for the same
reason. Therefore only Fig.5 and 6 contribute to the d* production.

 For the NN* intermediate state, only
  N*(1520 $IJ^p=\frac{1}{2}\frac{3}{2}^+$)
has been included. Because the results of GO\cite{go} show that the
contribution of NN*(1520) might be more important, the
NN*(1440) term will be left for further refinement. The $\Delta\Delta$ and
the D-wave components of deuteron are neglected temporary.

\section{Meson exchange current and cross section}
 The general electron scattering cross section formula of Donnelly and 
Raskin\cite{dr} will be used to calculate the inelastic $d^*$ production.
\begin{eqnarray}
\frac{d\sigma}{d\Omega}&=&\sigma_M(\nu_LR^L_{fi}+\nu_TR^T_{fi}+
\nu_{TT}R^{TT}_{fi}+\nu_{TL}R^{TL}_{fi})f^{-1}_{rec},
\end{eqnarray}
where $\sigma_M$ is the Mott scattering cross section,
\begin{eqnarray}
\sigma_M&=&\left(\frac{\alpha\cos\frac{\theta_e}{2}}{2\epsilon\sin^2\frac
{\theta_e}{2}}\right)^2,\\
\nu_L&=&\left(\frac{q^2}{\vec{q}^2}\right)^2,\\
\nu_T&=&-\frac{1}{2}\left(\frac{q^2}{\vec{q}^2}\right)+\tan^2\frac{\theta_e}{2},\\
\nu_{TT}&=&-\frac{1}{2}\left(\frac{q^2}{\vec{q}^2}\right),\\
\nu_{TL}&=&\frac{1}{\sqrt{2}}\left(\frac{q^2}{\vec{q}^2}\right)\sqrt{-\left(\frac{q^2}{\vec{q}^2}\right)+\tan^2\frac{\theta_e}{2}},\\
R^L_{fi}&=&\arrowvert\rho(\vec{q})_{fi}\arrowvert^2,\\
R^T_{fi}&=&\arrowvert J(\vec{q},+1)_{fi}\arrowvert^2+\arrowvert J(\vec{q},-1)_{fi}\arrowvert^2,\\
R^{TT}_{fi}&=&2Re\{J^*(\vec{q},+1)_{fi}J(\vec{q},-1)_{fi}\},\\
R^{TL}_{fi}&=&-2Re\{\rho^*(\vec{q})_{fi}(J(\vec{q},+1)_{fi}-J(\vec{q},-1)_{fi})\},\\
f_{rec}&=&1+\frac{2\epsilon}{M_T}\sin^2\frac{\theta_e}{2}.
\end{eqnarray}
$f_{rec}$ is the recoil correction. The four vector current $J^{\mu}(\vec{q})_{fi}$  is the Fourier transformed four vector
transition current of the target,
\begin{eqnarray}
J^{\mu}(\vec{q})_{fi}&=&\int d^3\vec{r}e^{i\vec{q}\cdot\vec{r}}\langle f\arrowvert  J^{\mu}(\vec{r})\arrowvert i\rangle,\\
\vec{J}(\vec{q})_{fi}&=&\sum_{m=0,\pm 1} J(\vec{q},m)\vec{e}(\vec{q};1,m),\\
\vec{e}(\vec{q};1,0)&=&\vec{u}_z,~
\vec{e}(\vec{q};1,\pm 1)=\mp\frac{1}{\sqrt{2}}(\vec{u}_x\pm i\vec{u}_y),\\
\vec{u}_x&=&-\frac{\vec{q}\times(\vec{k}\times\vec{k'})}{\arrowvert\vec{q}
\times(\vec{k}\times\vec{k'})\arrowvert},~
\vec{u}_y=\frac{\vec{k}\times\vec{k'}}{\arrowvert\vec{k}
\times\vec{k'}\arrowvert},~
\vec{u}_z=\frac{\vec{q}}{\arrowvert\vec{q}\arrowvert},\\
\rho(\vec{q})_{fi}&=&\frac{\arrowvert\vec{q}\arrowvert}{\omega}J(\vec{q},0)_{fi}.
\end{eqnarray}
The last Eq. is obtained from the four vector current conservation. $q$, $k$,
and $k^\prime$ are the four momentum transfer, the initial and final
momentum of the scattered electron depicted in Fig.7. $q=k-k^\prime$. 
$\theta_e$ is the electron scattering angle in the lab system. 
$M_T$ is the mass of the target, mass of deuteron in our case. 
$\epsilon$ ($\epsilon^\prime$) is 
the initial(final) energy of 
the scattered electron, and $\omega=\epsilon-\epsilon^\prime$.

\begin{figure}
\begin{center}
\rotatebox{180}{\psfig{figure=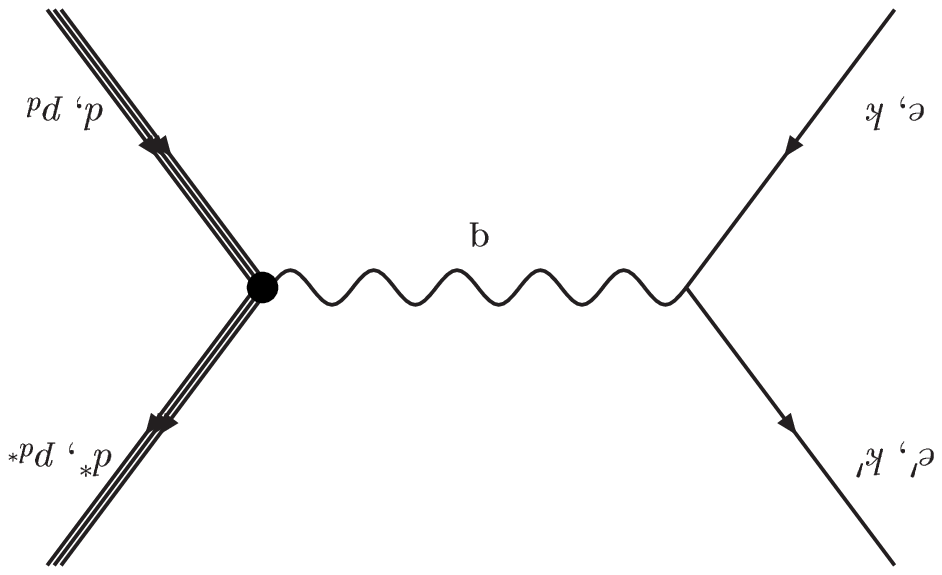,height=6cm}}
\caption{The electron-deuteron inelastic scattering}
\end{center}
\end{figure}

 For the unpolarized electron scattering, the cross terms $R^{TT}_{fi}$ and $R^{TL}_{fi}$ do not contribute. The angular momentum and parity conservation
($IJ^p=01^+$ for d and $03^+$ for $d^*$)restrict the multipole moments
further. Only the following 5 multipole moments contribute to the $ed\rightarrow ed^*$
process:
\begin{eqnarray}
F^C_2(\arrowvert\vec{q}\arrowvert)&=&\sqrt{\frac{4\pi}{3}}\langle 3\Arrowvert
 M_2(\arrowvert\vec{q}\arrowvert)\Arrowvert 1\rangle,\\
F^E_2(\arrowvert\vec{q}\arrowvert)&=&\sqrt{\frac{4\pi}{3}}\langle 3\Arrowvert
 T^E_2(\arrowvert\vec{q}\arrowvert)\Arrowvert 1\rangle,\\
F^M_3(\arrowvert\vec{q}\arrowvert)&=&\sqrt{\frac{4\pi}{3}}\langle 3\Arrowvert
 iT^M_3(\arrowvert\vec{q}\arrowvert)\Arrowvert 1\rangle,\\
F^C_4(\arrowvert\vec{q}\arrowvert)&=&\sqrt{\frac{4\pi}{3}}\langle 3\Arrowvert
 M_4(\arrowvert\vec{q}\arrowvert)\Arrowvert 1\rangle,\\
F^E_4(\arrowvert\vec{q}\arrowvert)&=&\sqrt{\frac{4\pi}{3}}\langle 3\Arrowvert
 T^E_4(\arrowvert\vec{q}\arrowvert)\Arrowvert 1\rangle.
\end{eqnarray} 
The multipole moments are defined through the following multipole expansion,
\begin{eqnarray}
J(\vec{q},0)_{fi}&=&\frac{\omega}{\arrowvert\vec{q}\arrowvert}\sqrt{4\pi}
\sum_{J\geq 0}\sqrt{2J+1}i^J\langle f\arrowvert M_{J0}(\arrowvert\vec{q}
\arrowvert)\arrowvert i\rangle,\\
J(\vec{q},\pm 1)_{fi}&=&-\sqrt{2\pi}\sum_{J\geq 0}\sqrt{2J+1}i^J\left(\langle f
\arrowvert T^e_{J,\pm 1}(\arrowvert\vec{q}\arrowvert)\arrowvert i\rangle\pm
\langle f\arrowvert T^m_{J,\pm 1}(\arrowvert\vec{q}\arrowvert)\arrowvert i\rangle\right),\\
M_{Jm}(\arrowvert\vec{q}\arrowvert)&=&\int d^3\vec{r}M_{Jm}(\arrowvert\vec{q}
\arrowvert\vec{r})\rho(\vec{r}),\\
T^E_{Jm}(\arrowvert\vec{q}\arrowvert)&=&\int d^3\vec{r}\frac{1}{\arrowvert
\vec{q}\arrowvert}(\bigtriangledown\times M_{JLm}(\arrowvert\vec{q}\arrowvert
\vec{r}))\cdot\vec{J}(\vec{r}),\\
T^M_{Jm}(\arrowvert\vec{q}\arrowvert)&=&\int d^3\vec{r}M_{JLm}(\arrowvert\vec{q}
\arrowvert\vec{r})\cdot\vec{J}(\vec{r}),\\
M_{Jm}(\arrowvert\vec{q}\arrowvert\vec{r})&=&j_J(\arrowvert\vec{q}\arrowvert r)
Y_{Jm}(\hat{r}),\\
M_{JLm}(\arrowvert\vec{q}\arrowvert\vec{r})&=&j_J(\arrowvert\vec{q}\arrowvert r)
Y_{JLm}(\hat{r}),\\
Y_{JLm}(\vec{r})&=&\left[Y_L(\hat{r})\otimes e(\vec{q};1)\right]_{JM}.
\end{eqnarray}

 The transition current resulted from Fig.5(a) is
\begin{eqnarray}
\rho\left(\vec{q}\right)&=&eF^N\left(q^2\right)\frac{M_N}{E\left(p\right)}
\frac{1}{p^0-E\left(p\right)+i\epsilon}
\left(\frac{f^*}{\mu}\right)^2\vec{S}^{\dagger}_1\cdot
\left(\vec{q}+\vec{q^\prime}
\right)
\vec{S}^{\dagger}_2\cdot\left(\vec{q}+\vec{q^\prime}\right)\\ \nonumber
&& F_{\pi}^2\left(\left(q+q^\prime\right)^2\right)
\frac{1}{\left(q+q^\prime\right)^2
-\mu^2+i\epsilon}
T^{\dagger}_1\cdot T^{\dagger}_2{\delta}^3\left(\vec{q}+\vec{p}_d
-\vec{p}_{d^*}\right),
\end{eqnarray}
\begin{eqnarray}
\vec{J}\left(\vec{q}\right)&=&e\left\{F^N\left(q^2\right)
\frac{\vec{p}_1+\vec{p}}{2M_N}+\frac{i}{2M_N}
\vec{\sigma}\times\vec{q}G^N_m\left(q^2\right)\right\}
\frac{M_N}{E\left(p\right)}\frac{1}{p^0-E\left(p\right)
+i\epsilon}\left(\frac{f^*}{\mu}\right)^2 \\ \nonumber
&&\vec{S}^{\dagger}_1\cdot\left(\vec{q}+\vec{q^\prime}\right)\vec{S}^{\dagger}_2\cdot
\left(\vec{q}+\vec{q^\prime}\right)F^2_{\pi}\left(\left(q+
q^\prime\right)^2\right) \frac{1}{\left(q+q^\prime\right)^2-\mu^2+i\epsilon}
 T^{\dagger}_1\cdot T^{\dagger}_2{\delta}^3(\vec{q}+\vec{p}_d-\vec{p}_{d^*}).
\end{eqnarray}

 A similar transition current resulted from Fig.6(a) is
\begin{eqnarray}
\rho(\vec{q})&=&\left(\tilde{g_{\gamma}}-\tilde{g_{\sigma}}\right)
\frac{\vec{S}^{\dagger}_1\cdot\vec{q}}{M_{N^*}-M_N}
\tilde{f}_{N^*\Delta\pi}
\frac{1}{\sqrt{S}-M_{N^*}
+i\frac{\Gamma\left(\sqrt{S}\right)}{2}}\left(-\frac{f^*}{\mu}\right)\\ \nonumber
&&F_{\pi}^2\left(\left(q+q^\prime\right)^2\right)
\frac{\vec{S}^{\dagger}_2\cdot
\left(\vec{q}+\vec{q^\prime}\right)}{\left(q+q^\prime\right)^2-\mu^2
+i\epsilon} T^{\dagger}_1\cdot T^{\dagger}_2{\delta}^3\left(\vec{q}+\vec{p}_d
-\vec{p}_{d^*}\right),
\end{eqnarray}
\begin{eqnarray}
\vec{J}\left(\vec{q}\right)&=&\left(\tilde{g}_{\gamma}-
\tilde{g}_{\sigma}\right)\vec{S}^{\dagger}_1
\tilde{f}_{N^*\Delta\pi}
\frac{1}{\sqrt{S}-
M_{N^*}+i\frac{\Gamma\left(\sqrt{S}\right)}{2}}\left(-
\frac{f^*}{\mu}\right)\\ \nonumber
&&F_{\pi}^2\left(\left(q+q^\prime\right)^2\right)\frac{\vec{S}^{\dagger}_2
\cdot\left(\vec{q}+\vec{q'}\right)}{\left(q+q^\prime\right)^2-\mu^2
+i\epsilon} T^{\dagger}_1\cdot  T^{\dagger}_2{\delta}^3
\left(\vec{q}+\vec{p}_d
-\vec{p}_{d^*}\right).
\end{eqnarray}

Fig.5(b) and Fig.6(b) will give similar transition currents.

 In obtaining these transition currents,the following effective Lagrangian have been used,
\begin{eqnarray}
L_{NN\gamma}&=& - e\overline{\Psi}_N\left(\gamma^\mu A_\mu -\frac{\chi_N}{2M_N}
\sigma^{\mu\nu}\partial_\nu A_\mu\right)\Psi_N \\
L_{N\Delta\pi}&=& - \frac{f^\ast}{\mu}\Psi^\dagger_\Delta S^\dagger_i\left(
\partial_i\phi^\lambda\right)T^{\lambda\dagger}\Psi_N  +  h.c. \\ 
L_{NN^*\gamma}&=& \overline{\Psi}_{N^{\ast}}\left[
\tilde{g_\gamma}\vec{S}^\dagger\vec{A}
-i\tilde{g_\sigma}(\vec{S}^\dagger\times\vec{\sigma})
\vec{A}\right]\Psi_N  +  h.c.\\
L_{N^*\Delta\pi}&=& i \overline{\Psi}_{N^{\ast}}
\tilde{f}_{N^{\ast} \Delta \pi}
\phi^\lambda T^{\lambda\dagger}
\Psi_\Delta  +  h.c.
\end{eqnarray}

 In these expressions $\Phi$, $\Psi_N$, $\Psi_{\Delta}$, $\Psi_{N^*}$, and
$A_{\mu}$ stand for the pion, nucleon, $\Delta$, $N^*(1520)$, and photon fields
respectively; $M_N$ and $\mu$ are the nucleon and pion masses; $\vec{\sigma}$
and $\vec{\tau}$ are the usual $1\over{2}$ spin and isospin Pauli operators; 
$\vec{S}^{\dagger}$ and $\vec{T}^{\dagger}$ are the transition spin and 
isospin operators from $1\over{2}$ to $3\over{2}$ with the normalization,
\begin{eqnarray}
\langle \frac{3}{2}, M  \left| S^{\dagger}_{\nu} \right|
\frac{1}{2}, m \rangle
= C \left( \frac{1}{2}, 1 ,\frac{3}{2} ; m, \nu, M \right)
\end{eqnarray}

 All the effective coupling constants, form factors, and the propagators
are taken from GO\cite{go}. They are copied here to make this report 
self contained and can be read without further check of many references.

$\chi^N = \left\{\begin{array}{c}
1.79\hspace{0.25cm}\text{proton} \\ -1.91 \hspace{0.25cm} \text{neutron}
\end{array} \right\}$ \\

$f^{\ast} = 2.13$\hspace{2cm}
$\tilde{f}_{N^{\ast}\Delta\pi} = 0.677$ \\

$\tilde{g}_\gamma = \left\{\begin{array}{c}
0.108\hspace{0.25cm} \text{proton} \\ -0.129\hspace{0.25cm}\text{neutron}
\end{array} \right\}$ \hspace{1.30cm}
$\tilde{g}_\sigma = \left\{\begin{array}{c}
-0.049 \hspace{0.25cm}\text{proton} \\ -0.0073\hspace{0.25cm} \text{neutron}
\end{array} \right\}$


 For the nucleon propagator,
only the positive energy part is retained where $E(p)=\sqrt{M_N^2+\vec{p}^2}$; 
for the $N^*$, the finite width $\Gamma$ has been included where 
\begin{eqnarray}
S=p^{02}-\vec{p}^2,\hspace{1cm}
\Gamma(\sqrt{S})=\Gamma(M_N^*)q_{cm}^5(\sqrt{S})/q_{cm}^5(M_N*).
\end{eqnarray} 
The form factors are taken from GO\cite{go} to keep the model consistent.
\begin{eqnarray}
F_\pi(q^2) = \frac{\Lambda_\pi^2 - m_\pi^2}{\Lambda_\pi^2-q^2};\hspace{1.5cm}
\, \Lambda_\pi\sim 1250\, MeV
\end{eqnarray}\\ 
Sachs's form factors are given by
\begin{equation}
G_M^N(q^2) = \frac{\mu_N}{(1 - \frac{q^2}{\Lambda^2})^2};\hspace{1.5cm}
\, G_E^N(q^2) = \frac{1}{(1 - \frac{q^2}{\Lambda^2})^2}
\end{equation}\\ 
with $\Lambda^2 = 0.71$ $GeV^2$; $\mu_p = 2.793$; $\mu_n = -1.913$.
\\
The relation
between $F_1^p(q^2)$   (Dirac's form factor) and $G_E^p(q^2)$ is :

\begin{equation}
F_1^p(q^2) = G_E^p(q^2)\frac{(1 -\frac{q^2\mu_p}{4m_N^2})}{(1 -
\frac{q^2}{4m_N^2})}
\end{equation}\\ 
and $F_1^n = 0.$

 To calculate the hadronic $d\rightarrow d^\ast$ transition current, a single
Gaussian wave function with a size parameter $b^\ast=0.7~fm$ is used to
approximate the $d^\ast$ internal motion. For the deuteron $d$ a three Gaussian
fitted to the ground state properties has been used, three size parameters
are $b=\left(4.0975,~1.8252,~0.8837\right)fm$ with normalization coefficient
$c=\left(0.31491,~0.49716,~0.36926\right)$.\cite{wong1} The $d^\ast$ mass is
taken to be 2.1 GeV. Due to the special spin(1 for d and 3 for $d^*$) and
orbital angular momentum(both 0) internal structure and the spin property
of the meson exchange current, only $F_2^C$, $F_2^E$ and $F_3^M$ three
transition form factors remain after the integration over the internal
spin and orbital variables of d and $d^*$. Fig.5 contributes one $F_2^C$,
one $F_3^M$ and two $F_2^E$(from convective and magnetic current
respectively), Fig.6 contributes one $F_2^C$ and one $F_2^E$ term. 

\section{results and discussions}
 The calculated $d^\ast$ electro production cross sections are shown in Fig.8
. The four curves correspond to four electron energies $0.8$, $1.0$, $1.2$
and $1.5~GeV$. The shape
of the differential cross section is dominated by the Mott cross section
but modulated by the inelastic transition form factors. The value around
30 degree in the laboratory frame is about 10 nb for 1 GeV electron.
In Fig.9, 10 and 11, the transition form factors $F_2^C$, $F_2^E$ 
and $F_3^M$ are
shown. The dashed curves correspond to the
contribution of $NN$ intermediate state while the full curves correspond to
the sum of contributions of two intermediate states. The $NN$ intermediate
state is the dominant one except 
at large angles, where the $NN^\ast$ contribution is more important. Fig.12
shows the differential cross section due to the NN intermediate state only
as well as a comparison to the full one.
 
 The produced $d^\ast$ will decay into $NN$ and $NN\pi$. The cross
section, 10 nb around 30 degree for 1 GeV electron, is about two order smaller than that of quasi
elastic $ed\rightarrow e^\prime NN$ and $\Delta$ resonance production
$ed\rightarrow e^\prime N\Delta
\rightarrow e^\prime NN\pi$ processes. Therefore the normal measurement of the
inelastic electron scattering can not find the signal of $d^\ast$ even it
is existed, i.e., the $d^\ast$ signal will be buried in the quasi elastic or 
$\Delta$ resonance background. A special kinematics and detector system
must be studied further both theoretically and experimentally in order to
pin down the weak signal of $d^\ast$ resonance from the strong background.

\begin{figure}
\begin{center}
\psfig{figure=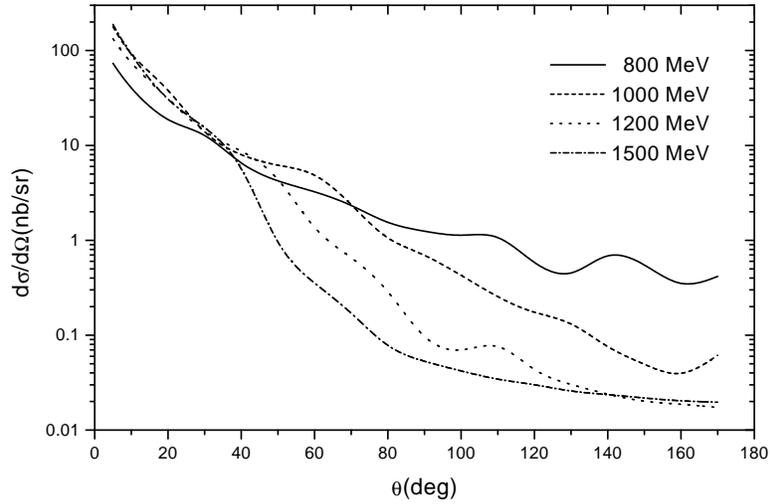,height=8cm}
\caption{$ed\rightarrow ed^\ast$ differential cross sections}
\end{center}
\end{figure}

\begin{figure}
\begin{center}
\psfig{figure=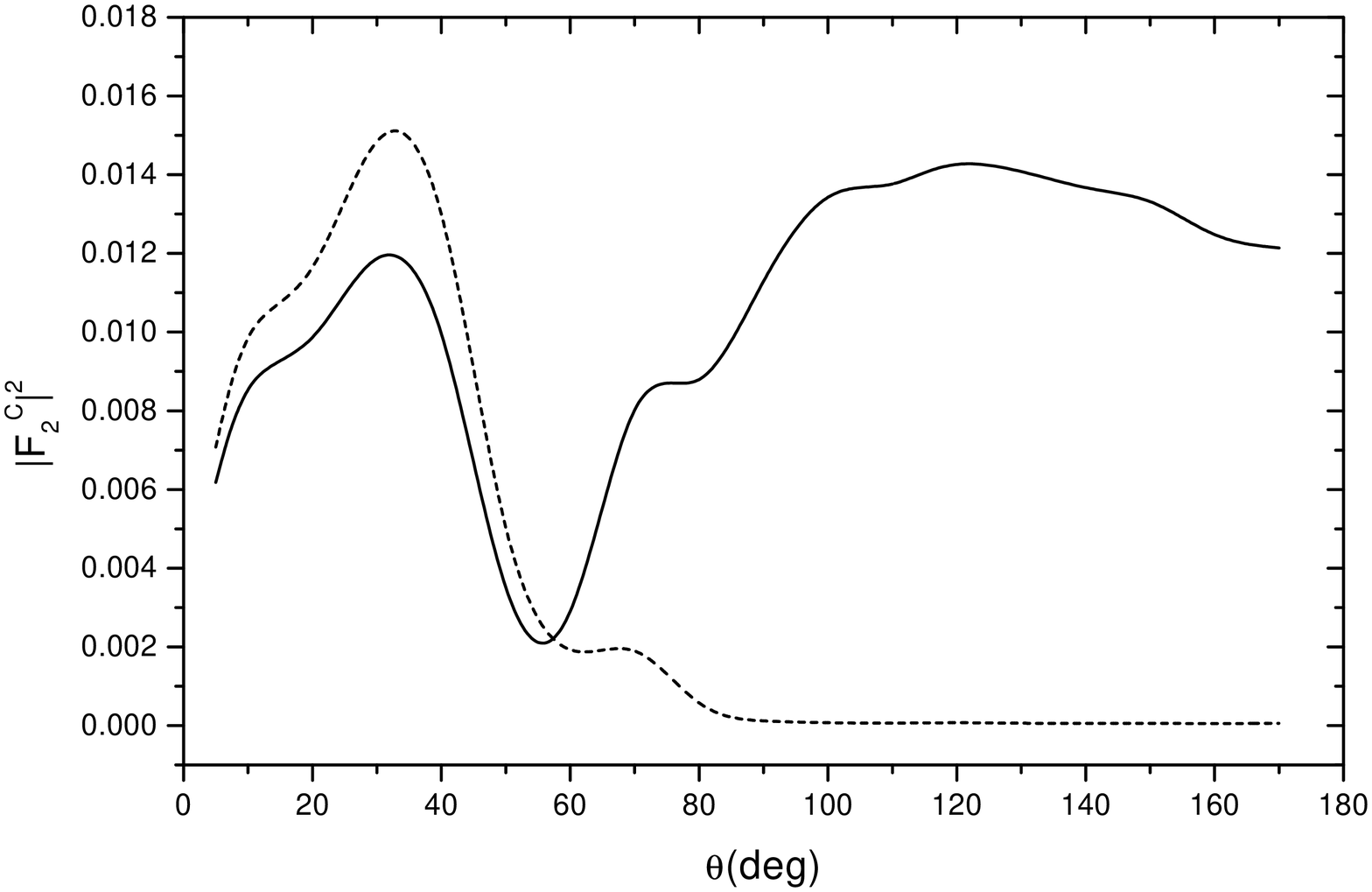,height=8cm}
\caption{$d\rightarrow d^\ast$ transition form factors: $\left|F^C_2\right|^2$,
$k_0= 1.5~GeV$, the dashed curve corresponds to 
the $NN$ intermediate state contribution, the full curve corresponds to the
sum of $NN$ and $NN^\ast$ intermediate state contributions.}

\psfig{figure=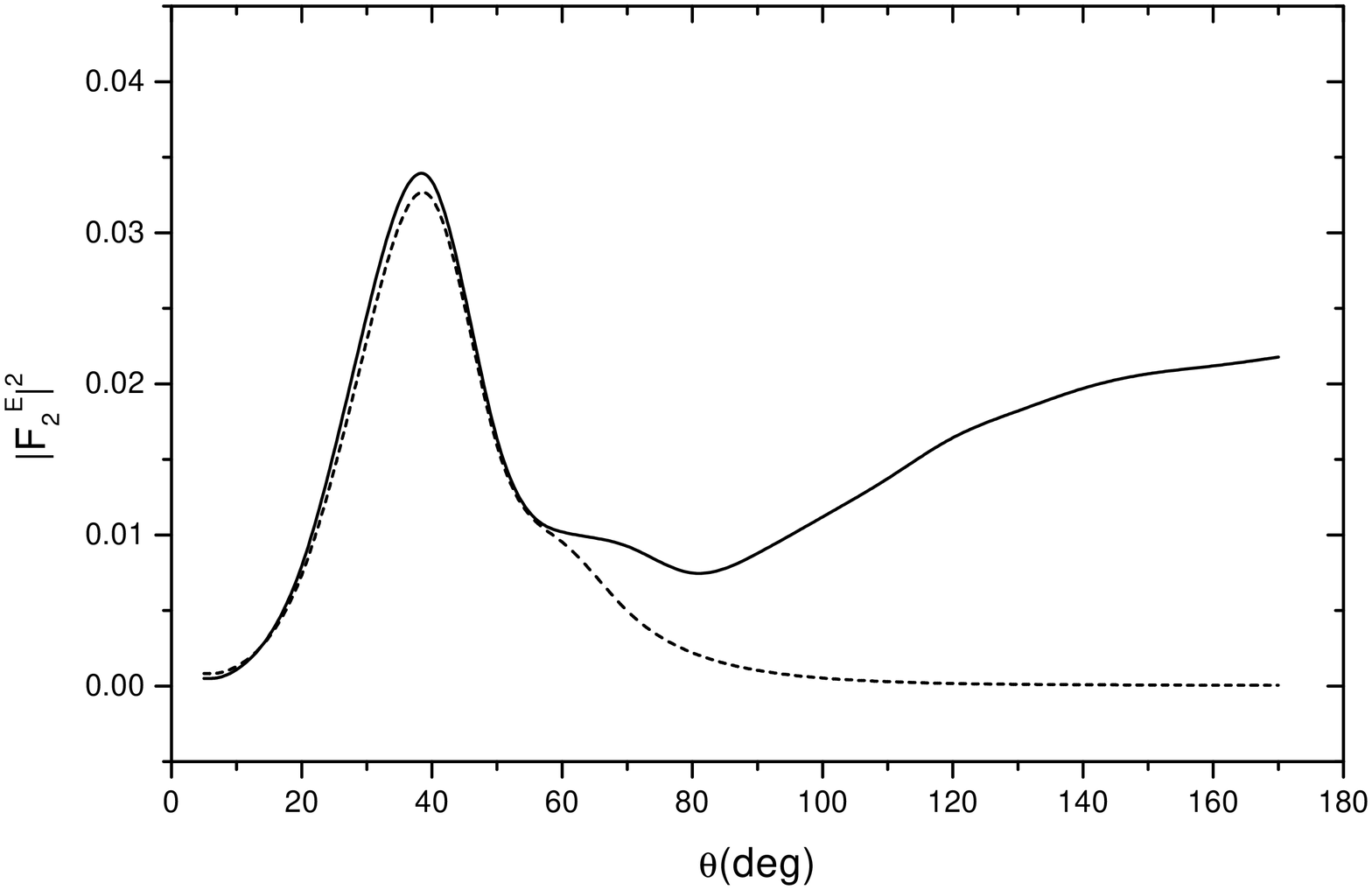,height=8cm}
\caption{Same as FIG. 9 for $\left|F^E_2\right|^2$}

\psfig{figure=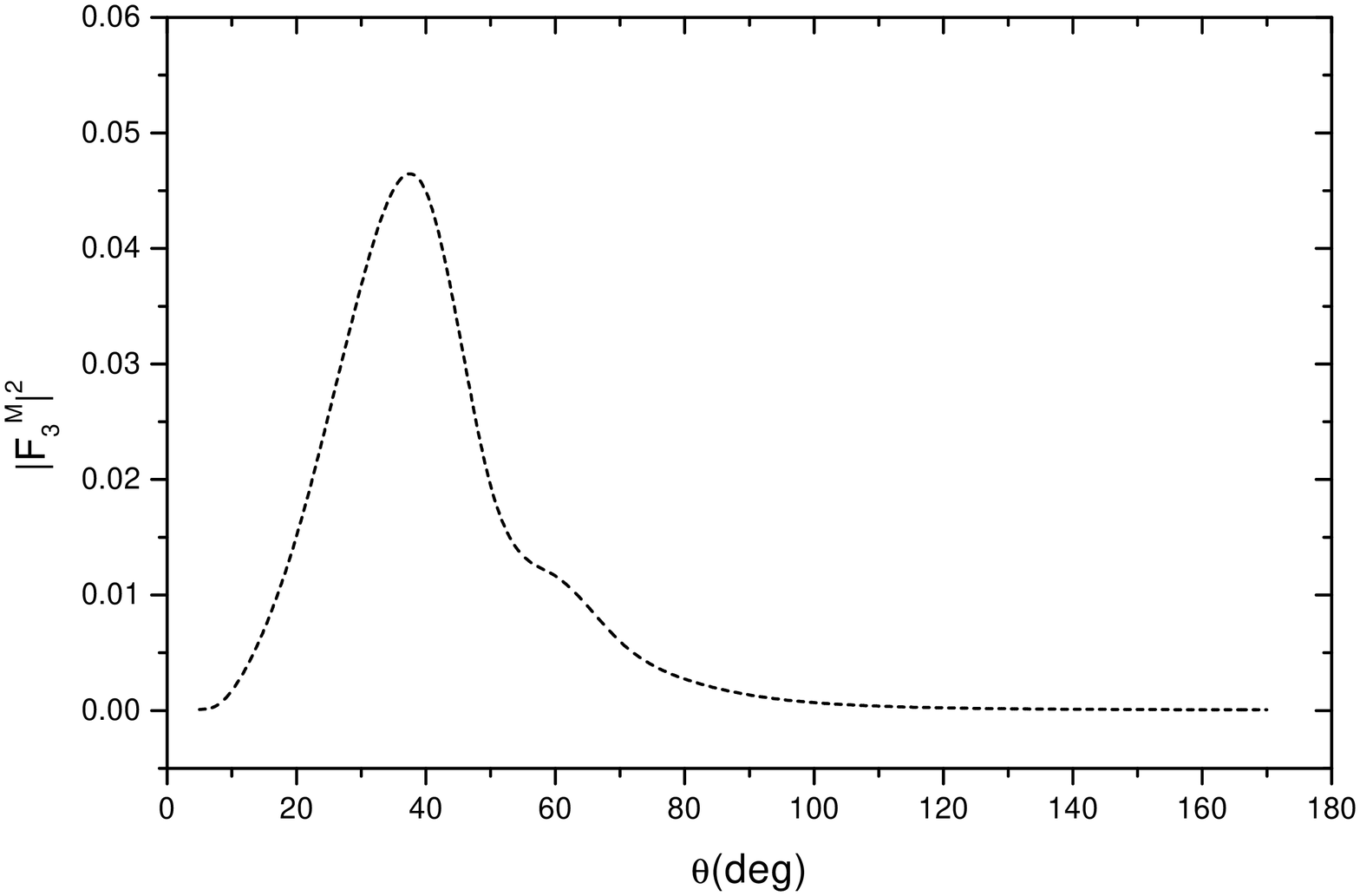,height=8cm}
\caption{Same as FIG. 9 for $\left|F^M_3\right|^2$}

\psfig{figure=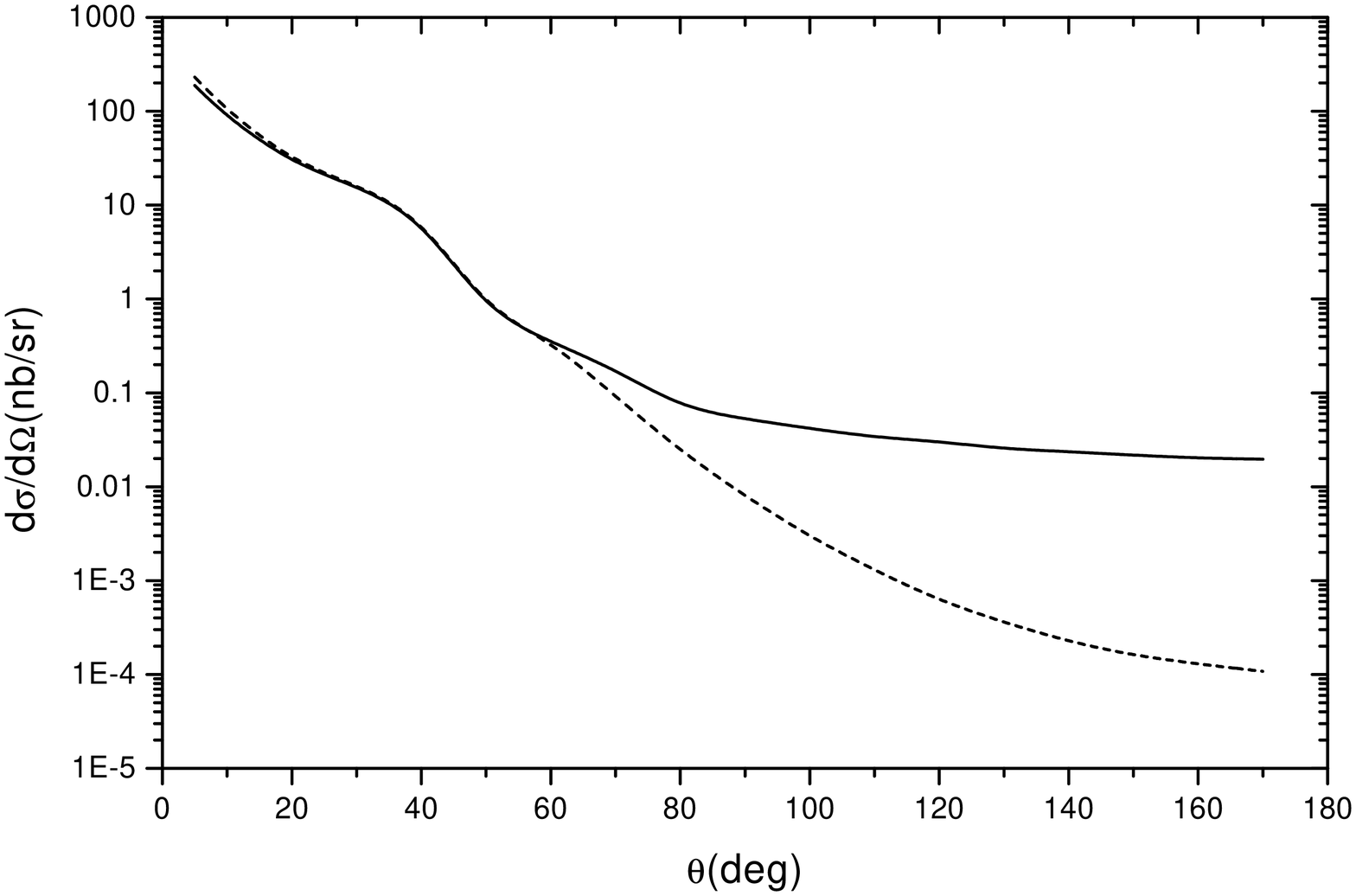,height=8cm}
\caption{Same as FIG. 9 for differential cross sections $d\sigma/d\Omega$}
\end{center}
\end{figure}

 In our model calculation, some simplifications have been assumed. The
D wave component of deuteron has been neglected. Even though it is a small
component, its contribution to the $d^\star$ production might not be small
enough to be neglected, because only a recoupling of spin and orbital 
angular momentum will be able to transit the deuteron $^3D_1$ into a $^3D_3$
NN component of $d^\ast$. The initial state correlation, i.e., the $\Delta-
\Delta$ component of deuteron is an even smaller component, but only a spin
recoupling is needed to transit it into $d^\ast$, therefore its contribution
should be checked as well. $d^\ast$ is a six quark state\cite{wang1} rather
than a $\Delta-\Delta$ bound state. To model it as a pure two $\Delta$ 
bound state and use a single Gaussian wave function to describe its
internal structure will overestimate the calculated cross section.
The contribution of $NN^*(1440)$ intermediate state should be added especially
in the large angle part. Due to these approximations the calculated cross
section is an estimate of the $d^*$ electro production. On the other hand all of the above mentioned corrections
are minor effects, the order(10 nb around 30 degree for 1 GeV electron)
obtained from this calculation might be stable against these fine tunes.
Further calculation is going on, especially a quark model calculation is also
doing to check if a single Gaussian approximation of the $d^\ast$ internal
wave function has overestimated the cross section.. The results will be reported later. 

The electro production of d' dibaryon is being measured. This calculation
can be extended to that process and will be a good check not only of the
electro production model used here but also of the d' dibaryon analysis itself.

Very helpful discussions with C.W .Wong, T. Goldman and Stan Yen are 
acknowledged. We are also greatly indebted to J.A. G\'omez Tejedor and E.
Oset for their helpful private communications.

This research is supported by NSF, SSTD and  the post Dr. foundation of SED of China. Part of the numerical calculation is done on the SGI Origin 2000
in the lab of computational condensed matter physics.


\begin{thebibliography}{99}

\bibitem{gold1} T. Goldman, K. Maltman, G.J. Stephenson,Jr., K.E. Schmidt,
    and F. Wang, Phys. Rev. {\bf C39}, 1889 (1989).
\bibitem{wong1} C. W. Wong, Phys. Rev. {\bf C57}, 1962 (1998).
\bibitem{wang1} F. Wang, G.H. Wu, L.J. Teng, and T. Goldman, Phys. Rev. 
    Lett. {\bf69}, 2901 (1992).
    F. Wang, J.L. Ping, G.H. Wu, L.J. Teng, and T. Goldman, Phys. Rev. {\bf C51},
    3411 (1995).
    T. Goldman, K. Maltman, G.T. Stephenson,Jr., J.L. Ping, and F. Wang, Mod.
    Phys. Lett. {\bf A13}, 59 (1998).
\bibitem{le} F. Lehar, in Baryons'98, eds. D.W. Menze and B.Ch. Metsch (World 
    Scientific, Singapore, 1999) p.622.
\bibitem{bi} R. Bilger, H. Clement and M. Schepkin, Phys. Rev. Lett. {\bf 71}, 42
    (1993).
\bibitem{ja} R.L. Jaffe, Phys. Rev. Lett. {\bf 38}, 195 (1977).   
\bibitem{yu} H. Yukawa, Proc. Phys. Math. Soc. Jpn. {\bf 17}, 48 (1935).
\bibitem{dema} M.C.M. Rentmeester, R.G.E. Timmermans, J.L. Friar and J.J.de Swart, 
    Phys. Rev. Lett. {\bf 82}, 4992 (1999) and references therein;
    C. Harzer, H. Muether and R. Machleidt, Phys. Lett. {\bf B459}, 1 (1999).
\bibitem{kf} T. Kamae and T. Fujita, Phys. Rev. Lett. {\bf 38}, 471 (1977).
\bibitem{wein} S. Weinberg, Physica {\bf 96A}, 327 (1979); Phys. Lett. {\bf B251},
    288 (1990); Nucl. Phys. {\bf B363}, 3 (1991).
\bibitem{ka} D.B. Kaplan, in Baryons'98, eds. D.W. Menze and B.Ch. Metsch (World
    Scientific, Singapore,1999) p.160.
\bibitem{gl} L.Ya. Glozman and D.O. Riska, Phys. Rep. {\bf 268}, 263 (1996);
    D.O. Riska and G.E. Brown, hep-ph/9902319.
\bibitem{mg} A.Manohar and H. Georgi, Nucl. Phys. {\bf B234}, 189 (1984).
\bibitem{fu} Y. Fujiwara, C. Nakamoto and Y. Suzuki, Phys. Rev. Lett. 
     {\bf 76}, 2242 (1996) and references there in. 
\bibitem{fae} M. Oka, K. Shimizu, and K. Yazaki, Phys. Lett. {\bf B130},
    365 (1983); Nucl. Phys. {\bf A464}, 700 (1987); M. Oka, Phys. Rev.
    {\bf D38}, 298 (1988); A. Faessler and U. Straub, Phys. Lett. 
    {\bf B183}, 10 (1987). 
\bibitem{rgg} A. De Rujula, H. Georgi and S.L. Glashow, Phys. Rev. {\bf D12}, 
     147 (1975).
\bibitem{is1} N. Isgur and  G. Karl, Phys. Rev. {\bf D18}, 4187 (1978); 
     {\bf D19}, 2653 (1979); {\bf D20}, 1191 (1979).
\bibitem{wong} C.W. Wong, Phys. Rep. {\bf 136}, 1 (1986) and references therein.
\bibitem{cjj} A. Chodos, R.L. Jaffe, K. Johnson, C.B. Thorn and V. Weisskopf, 
     Phys. Rev. {\bf D9}, 3471 (1974).
\bibitem{dsw} P.J.G. Mulders, A.T.M. Aerts, and J.J. de Swart, Phys. Rev. {\bf D17},
     260 (1978).
\bibitem{is2} N. Isgur, in Hadrons and Hadronic Mwtter,eds. D. Vautherin et al.
     (Plenum Press, New York, 1990) p.21
\bibitem{lo} R.L. Jaffe and F.E. Low, Phys. Rev. {\bf D19}, 2105 (1979);
     E.L. Lomon, Phys. Rev. {\bf D26}, 576 (1982);
     P. LaFrance and E.L. Lomon, Phys. Rev. {\bf D34}, 1341 (1986).
\bibitem{sim} Yu.A. Simonov, Phys. Lett. {\bf B107}, 1 (1981); Sov. J.
     Nucl. Phys. {\bf 36}, 422 (1982).
\bibitem{sw} T.H.R. Skyrme, Nucl. Phys. {\bf 31}, 556 (1962);
     E.Witten, Nucl. Phys. {\bf B160}, 57 (1979).
\bibitem{wa} T.S. Walhout and J. Wambach, Phys. Rev. Lett. {\bf 67}, 314 (1991);
     N.R. Walet and R.D. Amado, Phys. Rev. Lett. {\bf 68}, 3849 (1992).
\bibitem{jaf} R.L. Jaffe and C.L. Korpa, Nucl. Phys. {\bf B258)}, 468 (1985).    
\bibitem{wang2} G.H. Wu, L.J.Teng, J.L. Ping, F. Wang and T. Goldman, Mod. Phys.
     Lett. {\bf A10}, 1895 (1995); Phys. Rev. {\bf C53}, 1161 (1996).
\bibitem{gold2} T. Goldman, K. Maltman, G.J. Stephenson,Jr. and K.E. Schmidt,
     Nucl. Phys. {\bf A481}, 621 (1988); Phys. Lett. {\bf B324}, 1 (1994).
\bibitem{wu} J.L. Ping, F. Wang and T. Goldman, Nucl. Phys. {\bf A657}, 95 (1999);
     G.H. Wu, J.L. Ping, L.J. Teng, F. Wang and T. Goldman, LANL preprint 
     LA-UR-98-5841, hep-ph/9812079.    
\bibitem{yz} X.Q.Yuan et al., Commun. Theor. Phys. {\bf 32}, 169 (1999).
\bibitem{okay} M. Oka and K. Yazaki, in Quarks and Nuclei, ed. W. Weise(World
     Scientific, Singapore, 1984) p.489.
\bibitem{wal} R. Walet, Phys. Rev. {\bf C48}, 2222 (1993).     
\bibitem{nsf} US NSAC Long Range Plan URL: http://pubweb.bnl.gov/$\sim$nsac
\bibitem{cj} F.E. Close, R.L. Jaffe, R.G. Roberts and G.G. Ross, Phys. Rev.
     {\bf D31}, 1004 (1985).
\bibitem{go} J. A. G\'omez Tejedor and E. Oset,  Nucl. Phys. {\bf C571},
     667 (1994); {\bf A580}, 577 (1994); {\bf A600}, 413 (1996); 
     J.A. G\'omez Tejedor, E. Oset and H. Toki, Phys. Lett. {\bf B346}
     240 (1995); J. C. Nacher and E. Oset, nucl-th/9804006.
\bibitem{dr} T.W.Donnelly and A.S. Raskin, Ann. Phys. {\bf 169}, 287
       (1986).     
       
\end{thebibliography}
\end{document}